# SPREADING DEPOLARIZATION DETECTION IN ELECTROCORTICOGRAM SPECTROGRAM IMAGING BY DEEP LEARNING: IS IT JUST ABOUT DELTA BAND?

*Jeanne Boyer-Chammard[1*], Yinzhe Wu[1,2*], Chenyu Zhang[1,2],*
*Sharon Jewell[3], Anthony Strong[3], Guang Yang[1,2 †], Martyn Boutelle[1 †]*

[1]Department of Bioengineering, Imperial College London, London, U.K.
[2]I-X, Imperial College London, London, U.K.
[3]Department of Basic and Clinical Neuroscience, Institute of Psychiatry, Psychology and Neuroscience,
Academic Neuroscience Centre, King's College London, London, U.K.
[*]Co-first Author  [†]Co-last Author
{yinzhe.wu18, m.boutelle}@imperial.ac.uk

**ABSTRACT**

Prevention of secondary brain injury is a core aim of neurocritical care, with Spreading Depolarizations (SDs) recognized as a significant independent cause. SDs are typically monitored through invasive, high-frequency electrocorticography (ECoG); however, detection remains challenging due to signal artifacts that obscure critical SD-related electrophysiological changes, such as power attenuation and DC drifting. Recent studies suggest spectrogram analysis could improve SD detection; however, brain injury patients often show power reduction across all bands except delta, causing class imbalance. Previous methods focusing solely on delta mitigates imbalance but overlooks features in other frequencies, limiting detection performance. This study explores using multi-frequency spectrogram analysis, revealing that essential SD-related features span multiple frequency bands beyond the most active delta band. This study demonstrated that further integration of both alpha and delta bands could result in enhanced SD detection accuracy by a deep learning model.

***Index Terms—*** Electrocorticography (ECoG), Spectrogram Imaging, Deep Learning, Spreading Depolarization (SD), Traumatic Brain Injury (TBI)

## 1. INTRODUCTION

In neurocritical care, preventing secondary brain injury remains a critical focus, as it significantly impacts patient outcomes [1]. One prominent, independent contributor to secondary brain injury is the phenomenon of Spreading Depolarizations (SDs), which are pathological waves of near-complete depolarization that propagate from the initial focal lesion through the cerebral grey matter [2]. Monitoring SDs in clinical practice can use invasive electrocorticography (ECoG) with its high temporal resolution [3], [4] because of its intracranial access, which is essential for capturing the rapid electrophysiological changes associated with SD events.

Despite this, identifying SDs in ECoG recordings remains challenging [5]. The distinctive features of SDs, such as AC signal power attenuation, can be obscured by various signal artifacts when analyzing its full-band signal power [6] without resolution over the frequency axis [7], complicating reliable detection in clinical settings [8].

Recent research has explored the potential of spectrogram-based analyses [7], [9] to enhance SD detection beyond traditional electrophysiological metrics. However, brain injury patients often present with signal power attenuation across all frequency bands except for delta [10] (**Figure 1**). This can be explained by the preserved but abnormal synaptic activity due to structural damage or reduction of cerebral blood flow [10]. This frequency-specific attenuation contributes to class imbalance in spectrogram data, particularly when analyzing full-frequency bands. Previous approaches have attempted to circumvent this imbalance by restricting analysis to the delta band [7], [9], yet this simplification overlooks potentially valuable features across other frequency ranges, limiting detection performance.

**Our Contribution: (A)** This study examines spectral characteristics of SD-related signal changes beyond the delta band, showing that SD-induced power attenuation is detectable across multiple frequency bands. **(B)** Based on these findings, we propose the integration of spectrograms across diverse frequency bands to improve SD detection through a more complete spectral profile.

**Key Research Questions (RQ):**
**RQ1.** Are there useful features present in non-delta frequency bands (i.e., alpha, beta) for improved SD detection?
**RQ2.** Which frequency band best monitors SD occurrence?
**RQ3.** Can combining spectrograms from different frequency bands improve SD detection?

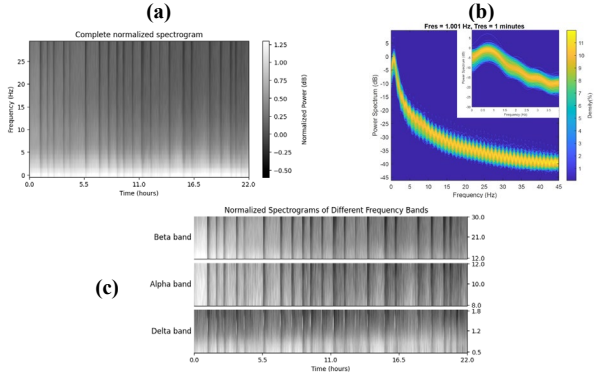

**Figure 1. (a)** Spectrogram of a typical brain injury patient experiencing spreading depolarization (SD). **(b)** Persistence Spectrum of another typical brain injury patient experiencing SD. (reproduced from [11] under its CC-BY license) Subfigures **(a)** and **(b)** illustrate a clear frequency-wise imbalance between the delta band (<4 Hz) and non-delta bands, highlighting typical characteristics in brain injury cases. **(c)** The spectrogram in (a) segmented by frequency bands—alpha, beta, and delta—and then signal power normalized for each band. **(c)** shows that segmented spectrograms for each frequency band, when each normalized, can highlight diverse features of spreading depolarization (SD), showing different relative contrasts at SD occurrences. These may be obscured in the original spectrogram due to low signal power intensity in the alpha (8-12 Hz) and beta (12-30 Hz) frequency bands.

## 2. METHOD

### 2.1. Dataset

Six brain injury patients were enrolled at King's College Hospital (London, UK) for this study. Inclusion criteria were a clinical decision for neurosurgical craniotomy and ages ranging from 18 to 80 years. Following sedation for ventilation and monitoring, all patients were in a pharmacologically induced coma. Given their comatose state at admission, written assent for participation was obtained from legally authorized representatives. Once the patients regained mental capacity, their own consents were acquired during follow-up. The study received approval from the KCH Research Ethics Committee (Cambridgeshire South; 05/MRE05/7) and the UC Institutional Review Board (2016-8153). The research was conducted in line with the Declaration of Helsinki.

An electrode strip was implanted under the dura, near the radiographically identified ischemic penumbra, following a craniotomy or decompressive craniectomy. This strip included six platinum electrodes from AdTech (Racine, WI, USA), each providing an ECoG signal channel. A patch electrode placed on the patient's neck served as the ground. After the procedure, the electrodes were connected to monitoring equipment in the intensive care unit (ICU) to begin data acquisition. ECoG data was captured using the Neuralink amplifier (USA) and recorded with LabChart software (ADInstruments, Sydney, Australia). At the end of the monitoring period, the strips were gently removed at the bedside without any complications related to their placement or removal. Full surgery details can be found in [4].

### 2.2 Data Pre-processing

To extract AC segment of the ECoG for SD detection, the signal is first filtered with a bandpass filter set between 0.5 and 45 Hz to isolate this segment. The filtered ECoG signal then undergoes a short-time Fourier transform (STFT) to produce a spectrogram with a time resolution of 1 minute. To assess **RQ1** and **RQ2**, the spectrogram is divided into **frequency bands** for specific analyses: **(1) alpha (8-12 Hz)** to assess **mid-frequency** responses to spreading depolarization (SD), **(2) beta (12-30 Hz)** to evaluate **higher frequency** responses to SD, and a **(3) restricted delta** range **(0.5–1.8 Hz)** to examine the **lower frequency** typically more active in brain injury patients. The restricted delta range is tailored to previous findings [7] to optimize SD detection in this frequency band. In line with clinical consensus [12] to further smooth out the signal artefacts, the leaky time integral of power at each minute is also computed to aid in SD detection; as per Wu et al. [9], combining the spectrogram and temporal power vector inputs can enhance detection accuracy. Before inputting to the model, each frequency band's spectrogram and power vector are normalized and segmented into overlapping 30-minute windows.

### 2.3 Model Design

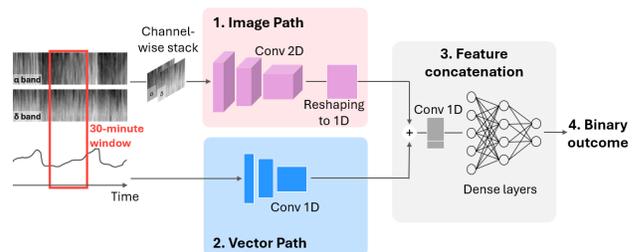

**Figure 2.** General framework for the identification of spreading depolarization (SD) using an ECoG tracing, analyzed across one or multiple frequency band spectrograms.

Initially, the signal is segmented using a 30-minute overlapping sliding window. Subsequently, the spectrogram of one or more frequency bands (alpha, beta, delta, or a channel-wise stacked combination of these) is fed into the **(1.) Image Path**. Meanwhile, the power vector—smoothed via a leaky time power integral—enters the **(2.) Vector Path**. The features from both paths are combined in the **(3.) Feature Concatenation** stage, leading to the final output of the **(4.) Binary Outcome** for SD Detection.

As introduced by Wu et al. [7] and in line with clinical consensus [12], joint inputs of spectrogram and temporal power vector could significantly improve the detection of SD over electrophysiology signals.

Similar to the Wu et al. [7], where they used a CNN chain to monitor SD in EEG, we adopted a two-way CNN backbone to jointly analyze 2D ECoG spectrogram and 1D temporal power vector for SD detection (**Figure 2**). The model is train by an Adam optimizer with binary cross-entropy (BCE) loss for 60 epochs.

To assess **RQ1** and **RQ2,** the image input to this chain is replaced by spectrograms of different frequency bands – specifically: **(1) alpha (8-12 Hz)** to assess **mid-frequency** responses to SD, **(2) beta (12-30 Hz)** to evaluate **higher frequency** responses to SD, and a **(3) restricted delta** range **(0.5–1.8 Hz)** to examine the **lower frequency** typically more active in brain injury patients. The restricted delta range is in line with previous findings [7] to optimize SD detection in this frequency band.

To further assess **RQ3**, the image input is further replaced with multiple spectrograms stacked channel-wise

**2.4 Evaluation Metrics**
For the binary outcome model outputs, accuracy, specificity, and sensitivity have been calculated for evaluation. In line with [7], a confidence score using a 30-minute summing window is also computed to visualize the detection results of the temporal sequence of SD detection within the context of temporal neighbors.

## 3. RESULT AND DISCUSSION

**3.1 RQ1: Are there useful features present in non-delta frequency bands (i.e., alpha, beta) for improved SD detection?**

As shown in **Table 1**, all assessed frequency bands (**alpha**, **beta**, **restricted delta**) achieved **high** performance in **accuracy** (>0.8) and **specificity** (>0.9), indicating that these bands contain features capable of accurately excluding SDs with minimal false positives.

**Table 1.** Average Accuracy, Sensitivity, and Specificity of the deep learning network results when solely input with spectrograms of alpha, beta, and restricted delta bands, respectively

| Frequency band | Accuracy(↑) | Sensitivity(↑) | Specificity(↑) |
|---|---|---|---|
| Restricted Delta (0.5-1.8Hz) | 0.8745 | **0.7130** | 0.9464 |
| Alpha (8-12 Hz) | **0.8853** | 0.7083 | **0.9630** |
| Beta (12-30Hz) | 0.8350 | 0.5768 | 0.9589 |

However, **sensitivity** analysis reveals a significantly **lower** sensitivity for the **beta** frequency band (12-30Hz) at 0.5768, compared to restricted delta at 0.7130 and alpha at 0.7083. This raises questions about the beta band's ability to show SD occurrence at higher frequencies, aligning with findings from [10] that suggest more silent brain activities in the beta frequency bands, indicating a possibly insignificant SD suppression contrast within this range.

Despite this, besides the restricted delta, which previous studies have shown capable of SD detection, the **alpha** band (8-12Hz) also demonstrated **comparable sensitivity** and even **higher specificity** and **accuracy**. This is notable since, although alpha range suppression is only moderately as indicated in **Figure 1(b)**, overall brain activity is reported to be greatly suppressed beyond the delta band [10]. This highlights the **alpha** range as a **potential additional diagnostic target** for SD monitoring in ECoG alongside the delta band.

**3.2 RQ2: Which frequency band best monitors SD occurrence?**

As shown in Table 1, the **alpha** band (8-12 Hz) achieved the **highest accuracy** (0.8853) and **specificity** (0.9630) across all frequency bands, both statistically significant compared to the runner-up restricted delta band ($p < 0.01$).

Although the restricted delta band exhibited the highest **sensitivity** (0.7130) relative to the alpha band (0.7083), the difference was **not statistically significant**. This leads to the conclusion that the **alpha band** secures the **highest accuracy and specificity** with a **sensitivity comparable** to the best performing restricted delta band, making it the optimal frequency band for SD monitoring. This finding is further supported by the plotted validation loss during training and the visualized confidence scores (**Figure 3**).

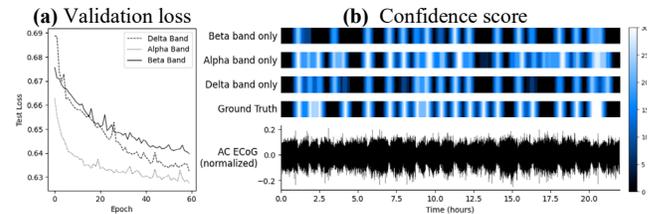

**Figure 3. (a)** Validation BCE loss over epochs, showing that the alpha band maintains significantly lower loss across all epochs. Initially, the delta and beta bands display similar losses, but later in training, the delta band diverges, achieving lower validation BCE loss than the beta band. **(b)** Confidence score visualized with a 30-minute summing window, following the method in [7], indicating significantly improved results with alpha band detection.

This result is surprising given that the restricted delta band has the highest signal power (**Figure 1(b)**) and represents the most active electrophysiological range under brain injury [10]. This may suggest that the relative contrast of SD suppression may become more pronounced in the alpha range, which supports more complex brain activities.

**3.3 RQ3: Can combining spectrograms from different frequency bands improve SD detection?**

Based on the conclusions from RQ1 and RQ2, we evaluated whether combining the alpha and delta bands could further enhance detection outcomes. By concatenating the alpha and delta bands channel-wise, the final model demonstrated a **dramatically higher accuracy (0.9193**, $p < 0.05$ compared to restricted delta alone), a **modest increase** in **sensitivity** (**0.8223**, $p = 0.06$ compared to restricted delta alone), and **comparable specificity (0.9635**, $p > 0.05$ compared to restricted delta alone).

## 3.4 Significance of studying SD signature across different frequency bands

Previous work by Bastany et al. [9] and Wu et al. [7] demonstrated the relevance of leveraging the ECoG spectrogram for SD identification. In both cases, the spectrogram was restricted to the delta band, i.e., 0.5–4 Hz, as it exhibits the highest power [7]. However, according to Nasretdinov et al. [13], changes in activity following an SD range from depression to boom. The variability is mostly found in the delta band, whereas higher frequency bands exhibit a stereotypical depression of activity [13]. Thus, restricting the analysis to the delta band could result in significant undersensitivity.

Since power is lower in higher frequency bands compared to the delta band, normalizing the entire spectrogram together can lead to a loss of information. Our approach involved computing frequency band-restricted spectrograms for the delta and alpha bands and normalizing them separately. This enhances visualization of the relative contrast resulting from the AC power suppression associated with SD occurrence, aiding in their detection.

These findings open new avenues for SD monitoring strategies, which could be retrospectively informative to the clinical practice and advocates for an expanded spectral approach that includes alpha as an additional frequency band to improve SD detection accuracy and reliability in clinical settings. Future work could refine multi-band integration to further optimize SD monitoring frameworks in neurocritical care environments.

## 4. CONCLUSION

This study examined the spectral characteristics of SD in ECoG data across various frequency bands to enhance SD detection in neurocritical care. While the delta band has traditionally been emphasized, our findings show that the alpha band also yields high accuracy and specificity with comparable sensitivity, suggesting it as an optimal frequency for SD monitoring. Moreover, combining alpha and delta bands significantly enhances detection accuracy, supporting a more comprehensive, multi-band approach to SD monitoring. This study highlights the potential of multi-band spectrogram analysis to improve SD detection in neurocritical care. It also advocates for broader spectral integration in SD detection frameworks, potentially informing future neurocritical care practices.

## 13. ACKNOWLEDGMENTS

This work is funded in part by Imperial College London President's PhD Scholarship. G. Yang's work is supported by the UKRI Future Leaders Fellowship (MR/V023799/1).